\documentclass[12pt]{iopart}
\usepackage{iopams}  
\usepackage[all,2cell,dvips]{xy}
\def\oint{{\relax \int\kern -1. em O}}

\def\ba{\begin{eqnarray}}
\def\ea{\end{eqnarray}}
\newcommand{\eq}{\begin{equation}}
\newcommand{\feq}{\begin{equation}}
\newcommand{\bea}{\begin{eqnarray}}
\newcommand{\eea}{\end{eqnarray}}

\font\tengoth=eufm10 \font\sevengoth=eufm7 \font\fivegoth=eufm5
\newfam\gothfam
\textfont\gothfam=\tengoth \scriptfont\gothfam=\sevengoth
  \scriptscriptfont\gothfam=\fivegoth
\def\<#1>{\langle#1\rangle}

\def\be{\begin{equation}}
\def\ee{\end{equation}}

\newenvironment{tabla}{\fontsize{8}{17}\selectfont}{}
\def\bt{\begin{tabla}}
\def\et{\end{tabla}}
\def\bt{{\bf t}}

\begin{document}
% Journal identifier can be put here if required, e.g.
\jl{2}

\title[Octonions and $M$-theory]
{Octonions and $M$-theory}

\author{Luis J. Boya\footnote[3]{Presented at the the 24th International
 Colloquium on Group-Theoretical Methods in Physics. Paris, July 15-22, 2.002}}

\address{Departamento de F\'{\i}sica Te\'orica. Facultad de Ciencias. \\
Universidad de Zaragoza, E-50009, Zaragoza, Spain.\\ email: {luisjo@posta.unizar.es}}

\begin{abstract}
        We explain how structures related to octonions are ubiquitous in 
$M$-theory. All the exceptional Lie groups, and the projective Cayley 
line and plane
appear in $M$-theory. Exceptional $G_2$-holonomy manifolds show up as
compactifying spaces, and are related to the $M$2 Brane and 3-form. We review
this evidence, which comes from the initial 11-dim structures. 
Relations between
these objects are stressed, when extant and understood. We argue for the
necessity of a better understanding of the role of the octonions themselves (in
particular non-associativity) in $M$-theory.
\end{abstract}


%\submitted

%\maketitle

\section{Introduction}

If the current $M$-theory is a {\sl unique
theory}, one should expect it to make use of singular, non-generic
mathematical structures. Now it is known that many of the special
objects in mathematics are related to octonions [1], and therefore
it is not surprising that this putative {\sl theory-of-everything} should
display geometric and algebraic structures derived from this
unique non-associative division algebra.

        Special algebraic objects related to octonions are the five
exceptional simple Lie groups $G_2$, $F_4$, $E_6$, $E_7$, and $E_8$. 
Some sporadic
(finite simple) groups seem also related to octonions (see e.g.
Thomson's $E_2(3)$ [2]).The use of them in future directions of the
theory is not to be discarded. The Cayley plane and its predecessor
the octonionic projective line on one hand, and the geometries
associated to the Magic Square [3] on the other, stand as
fundamental octonionic geometries.

        Very recently, the newly discovered (1996) compact
manifolds of $G_2$-holonomy (a case of Joyce manifolds [4]) might
play a fundamental role also in $M$-theory.

Also, the four coincidences in the list of simple Lie groups
$$
\begin{array}{cccc}
A_1=B_1=C_1&              B_2=C_2  &                 A_3=D_3 &      D_2=A_1\times A_1\cr
SU(2)={\rm Spin}(3)=Sp(1) &  {\rm Spin}(5)=Sp(2)    &     SU(4)={\rm Spin}(6)&
{\rm Spin}(4)=SU(2)\times SU(2)
\end{array}
$$
are related to some irreps.  of the exceptional groups (Adams [5]);
they also appear in physics and in compactifications in $M$-theory
disguised in various forms [6].

        What we lack at the moment is an understanding of the role
of the octonions themselves in $M$-theory; for a previous discussion
of the history of the role of octonions in physics see [7]

\section{Projective lines and planes}

The first apparition of
octonions in $M$-theory is as the fourth  ladder in the ``Brane Scan''
of Townsend (1987) [8]. The four lists of classical supersymmetric
$p$-Branes including instantons, embedded in $D$-dimensional 
space(-time) correspond precisely to the four division algebras $\mathbb{R}, 
\mathbb{C}, \mathbb{H}, \mathbb{O}$.

\bigskip

Real ladder: From $D=1$\  $p=-1$ instanton (kink) to $D=4$\  $p=2$ membrane
\centerline{(domain wall; codimension one)}

Complex ladder: From $D=2$\  $p=-1$ instanton (vortex) to $D=6$\  $p=3$
``universe''

\centerline{(codimension two)}

$\mathbb{H}$: Quaternionic ladder: From $D=4$\  $p=-1$ instanton to $D=10$\ 
        $p=5$ brane 

\centerline{(codimension four)}

Octonionic ladder: From $D=8$\  $p=-1$, Fubini-Nicolai instanton [9] to $D=11$
$ p=2$
$M2$ membrane

\centerline{ (codimension eight)}

\medskip

\centerline{{\underline{Table I}}.- The Brane Scan (Townsend [8])}

\bigskip

        These four ladders can be thought of as ``oxidation'' of the
corresponding instanton, in its turn associated to the fundamental
line bundle for the projective spaces $\mathbb{KP}^1$, $\mathbb{K}=\mathbb{R}, \mathbb{C}, \mathbb{H}, \mathbb{O}$ [10]. The
four are supersymmetrizable and are linked as the four Hopf
bundles
$$
\xymatrix{
&&&&&&&&{\rm Group}&{\rm Susy}\\
\alpha:&S^0\ar[r]&S^1\ar@{=}[d]\ar[r]&S^1\ar@{=}[r]&\mathbb{RP}^1&&&&O(1)& 1+1\\
\beta:&&S^1\ar[r]&S^3\ar[r]\ar@{=}[d]&S^2\ar@{=}[r]&\mathbb{CP}^1&&&U(1)&2+2\\
\gamma:&&&S^3\ar[r]&S^7\ar[r]\ar@{=}[d]&S^4\ar@{=}[r]&\mathbb{HP}^1&&Sp(1)&4+4\\
\delta:&&&&S^7\ar[r]&S^{15}\ar[r]&S^8\ar@{=}[r]&\mathbb{OP}^1&{\rm Spin}(8)
&8+8}
$$
\medskip

\centerline{{\underline{ Table II}}.- Elementary Solitons and Projective Lines [10]}

\bigskip

        The relation of the series with the numbers $\mathbb{R},\ldots,\mathbb{O}$ can be
made more precise [11], although the octonion case is more obscure
[12]. The elementary objects (e.g. strings for $p =1$) are ``thin'' (i.e.,
strictly $p$ space dimensions) but the soliton has a width; as
emphasized by Townsend ([8], [13]) the elementary ``thin''
membrane has a continuous excitation spectrum, but the
membrane really grows a ``core'' due to gravitation.

        Notice in Table I the $\mathbb{O}$-series appears as a ``second coming'' of
the Real ladder: both end up in a membrane. Quantization seems to
select only the $\mathbb{O}$  ladder.

        The relation of the Cayley plane $\mathbb{OP}^2$ (R. Moufang, 1933) to
the 11 $D$ Sugra ``corner'' of $M$-theory is more recent: it comes from a
nice paper of Ramond [14]. The Cayley plane is the 16 $D$ rank one
symmetric space (compact form)
$$
                F_4/B_4 :   1\to {\rm Spin}(9)\to F_{4(-52)}\to \mathbb{OP}^2\to 1\qquad 52=36+16
$$

        Now the ratio $\rho$ of the orders of the Weyl's groups is 3, as
$\#$ Weyl($F_4$)=1152 and Weyl($B_4$)= $\mathbb{Z}_2\,^4\odot S_4$, of 
order 384. But when a
pair $H$, $H\subset G$ of semisimple Lie groups have the same rank, an
important construction of Konstant {\sl et al.} [15] generates, for each
irrep of $G$, $\rho$ irreps of $H$, where $\rho= [{\rm Weyl}(G):{\rm Weyl}(H)]$ ($= 3$ in our
case). In particular the Id(-entity) irrep of $F_4$ generates
$$
    {\rm Id}(F_4)\to + 44({\rm graviton}) - 128({\rm gravitino}) + 84(3-{\rm form})\quad   {\rm of\  Spin}(9)
$$
i.e. precisely the Spin(9) content of 11 $D$ Sugra, where Spin(9) is the
little group in the light cone!

\section{The $E$-series}

 The evidence for the $E_1-E_{10}$ series in
the descent of Sugra 11 $D$ down to $D=1$ is well-known, and first
stated by Julia in 1982 [16]. Compactifying 11 $D$ Sugra from the
original 11 $D$ to 3 $D$, all of the $E$ series of {\sl split} forms appear
successively; the moduli spaces of scalar fields are the homogeneous
spaces. We recall only the non-compact/compact scalars:
\begin{center}
\begin{tabular}{clr}
&&                                                            \qquad          scalars\\
5 $D$ \qquad&                     $ E(6, +6)/Sp(4)$ &                                   42\\
4 $D$ \qquad          &            $E(7, +7)/SU(8)$ &                             35+35\\
3 $D$   \qquad       &            $ E(8, +8)/SO(16)$&                                128
\end{tabular}
\end{center}

\bigskip

On the other hand, the Heterotic Exceptional string, of
course, makes an important and direct use of $E_8\times E_8$, and its descent
from 11 $D$ $M$-theory has been clarified in the fundamental work of
Witten and Horawa [17], through compactification in a segment.
Finally, we recall that $E_6$, which appears naturally in some H-E
string compactifications, is a strong candidate for Grand Unified Theories.

\section{Manifolds of $G_2$ holonomy}

In the ``old''
superstring compactification $10\ D\to  4\ D$,  Ricci-flat Calabi-Yau 3-folds 
were the objects of course. In $M$-theory with 11 $D$, their place
is taken by manifolds with exceptional holonomy (of the Berger
1955 list  [18]). In particular, $G_2$ compact holonomy manifolds are
still Ricci flat and conserve 32/8=4 supercharges, that is, $N=1$ Susy
in $4\ D$ as we want. These beasts are fairly new even for the
mathematicians  (Joyce manifolds, 1996 [4]). From the reduction
of the tangent bundle of a compactifying space $K_7$, $M$-theory(11)$\to 
K_7\,+\,$Minkowski

$$
\xymatrix{
G_2\ar[d]& & \\
SO(7)\ar[d]\ar[r]&P\ar[r] &K_7\\
\mathbb{RP}^7\ar[r]&E\ar[r]&K_7
}
$$
we see that $G_2$ holonomy involves some torsion properties in $K_7$, a
hot topic today. The structure is given in two diagrams
$$
\xymatrix{
SU(3)\ar[r]\ar[d] &
SU(4)={{\rm Spin}}(6)\ar[r]\ar[d]&
S^7\ar@{=}[d]&
\qquad &
{}&
\mathbb{Z}/2\ar@{=}[r]\ar[d]
& \mathbb{Z}/2\ar[d]
\\                  
G_2\ar[r]\ar[d]
&
{\rm Spin}(7)\ar[r]\ar[d]
&
S^7&
\qquad &
G_2\ar[r]\ar@{=}[d]
& 
{\rm Spin}(7)\ar[r]\ar[d]
&S^7\ar[d]
\\
S^6\ar@{=}[r]&
S^6&
&
\qquad &
G_2\ar[r]&
SO(7)\ar[r]&
\mathbb{RP}^7
 }
$$

They come from the curious fact that $G_2$ is subgroup of
Spin(7) (Adams [5]) and of $SO(7)$ (from the 7 irrep). For a review
of the involved physics see [19]. The reason for the descent $SO(7)$
to $G_2$ is because the later conserves a 3-form, related to the
octonionic product [20], and very likely to the 3-form extant in $M$-theory
 in the 11 $D$ Sugra limit (see before). This is the only
rationale, so far, for the presence of the octonions {\sl themselves} in $M$-theory; we hope these matters will be clarified soon.

For alternative and/or complementary views of the items
exposed here see [21], [22].

\end{document}